\documentclass{article}
\usepackage{graphicx,psfrag,verbatim}
\usepackage{latexsym}
\usepackage{amssymb}
\usepackage{amsfonts}
\usepackage{amsmath}
\usepackage{graphicx}
\usepackage{verbatim}
\numberwithin{equation}{section}

\begin{document}

\markboth{Marcelo L. Costa, Amilcar R. Queiroz \& Ademir E.
Santana } {Non-commutative Thermofield Dynamics}

%
%

\title{Noncommutative Thermofield Dynamics}

\author{Marcelo L. Costa$^{a,b}$\footnote{leineker@fis.unb.br}, Amilcar R.
Queiroz$^{b,c}$\footnote{amilcarq@unb.br} \ and \ Ademir E.
Santana$^b$\footnote{asantana@fis.unb.br} 
\vspace*{0.5cm}\\ 
$^a$ N\'ucleo de F\'isica, Universidade Federal do Tocantins,\\
 Palmas, TO, 77001-090, Brazil\\
$^b$ Instituto de F\'{\i}sica, Universidade de Bras\'{\i}lia\\
Bras\'{\i}lia, DF 70910.900, Brazil
\\
$^c$ International Center for Condensed Matter Physics,\\
Universidade de Brasília
\\
Bras\'{\i}lia, DF, 70910.900, Brazil}

\maketitle


\begin{abstract}
The real-time operator formalism for thermal quantum field
theories,  thermofield dynamics, is formulated in terms of a
path-integral approach in non-commutative spaces. As an
application, the two-point function for a  thermal
non-commutative $\lambda \phi^4$ theory is derived at the one-loop
level. The effect of temperature and the non-commutative
parameter, competing with one another, is analyzed.
\end{abstract}



\section{Introduction}

The quantum field theory formulated in a non-commutative
space-time has attracted attention over the last decades due to a
variety of motives. In particular, it has been associated with the
behavior of non-abelian gauge fields and with phenomenological
effects in condensed matter physics, including aspects of phase
transitions~\cite{suss,matt12,matt1,matt222,matt22,matt2,matt3,ball114}.
In addition, there are some results coming from string theory. In
this case,  Connes, Douglas e Schwarz \cite{schwarz} showed that
an M-theory can be equivalent to a supersymmetric Yang-Mills field
in a
non-commutative torus; a result explored and expanded by Seiberg e Witten%
\cite{seib}.

A key ingredient of non-commutative quantum field theories is the
mixing of ultraviolet  and infrared divergence (UV-IR mixing)
arising simultaneously in perturbation schemes. Such a divergence
is a signature of non-commutativity due to the appearance of
non-planar diagrams; and a way to regularize the UV divergences is
by requiring that the external momenta be away from the IR regime.
The UV-IR mixing is relevant for applications of non-commutativity
to condensed matter systems, including the finite temperature
cases, as in the quantum hall effect~\cite{suss}. For instance, a
way for testing the non-commutativity in space coordinates is to
consider the limit of high temperature~\cite{ar}. These elements
have motivated several studies to take into account temperature,
and all them are based on the Matsubara
formalism~\cite{fischler,bal1,bal11,bal112,bal113}.

In thermal field theory one finds two approaches  that are
introduced by different choices of  complex-time contours. The
imaginary-time (Matsubara) formalism~\cite{matsubara} takes the
Boltzmann factor, $ \exp(-\beta H)$, where $\beta $ is the inverse
of temperature $T$ ($\beta =1/T$), under a Wick rotation in the
time evolution, such that the time $t$ is mapped into an imaginary
time: $t\rightarrow t=i\tau ,$ with $0\leq \tau \leq \beta .$ The
other formalism is called real-time and, since there is no
imaginary time, it can be applied to non-equilibrium problems. The
first real-time approach is due to Schwinger~\cite{schwinger} and
Keldysh~\cite{keldysh} including a functional development. Another
method considering real-time is called thermofield dynamics
(TFD)~\cite{ume1,kha1}, which is based on the notion of linear
space and representation theory. Both formalisms,
Keldysh-Schwinger and TFD, are essentially the same in terms of
results for equilibrium thermodynamics~\cite{chu1}.

TFD was proposed by Takahashi and Umezawa in order to overcome
difficulties with the imaginary-time~\cite{3ume1}. The basic
elements they used were a doubling of the Hilbert space and  a
Bogoliubov transformation. The latter is a rotation, associated
with the doubling, that leads to thermalization. This formalism
has been developed for practical purposes. These include
perturbative schemes and Feynman rules that follow in parallel
with the zero-temperature quantum field theory~\cite{ume1,3ume2}
and applications to superconductivity~\cite{3ume3}, magnetic
systems like ferromagnets and paramagnets~\cite{3ume22}, quantum
optics~\cite{3bar,3rev1,3rev2,3cha1,3cha2}, transport
phenomena~\cite{3rev5}, d-branes~\cite{3ale1,3ale2} and particle
physics~\cite{3ume2}.

Formally the real-time theory can be studied by using TFD in the
context of c*-algebras~\cite{3oji1} and symmetry
groups~\cite{kha1,3oji2,ade44}. In this analysis the structure of
the Tomita-Takesaki (standard) representation~\cite{3tom2,brate}
is used and, in particular, the physical meaning of the doubling
is fully identified with the notion of thermo-Lie
groups~\cite{3ak1,kha1000}. Along these lines, there are several
possibilities to explore thermal effects. As an example, the
kinetic theory has been formulated by analyzing representations of
symmetry groups~\cite{kha1} and elements of the q-group have been
considered, where the effect of temperature is related to a
deformation in the Weyl-Heisenberg algebra~\cite{3ak3,3ak4}.

One interesting aspect coming naturally from the algebraic
structure of TFD is that the propagator is written in two pieces:
one describes the $T=0$ theory, while the other gives rise to the
temperature effect. This is not the case of the imaginary time
formalism, where the propagator reduces to that of a 3-dimensional
Euclidian theory in the momentum space, involving an infinite sum
of the Matsubara frequencies: $w_{n}=2\pi n/\beta $ in the case of
bosons and $w_{n}=\pi (2n+1)/\beta $ for fermions. TFD should then
be of interest when the effect of temperature is competing with
another parameter, as in a non-commutative approach. In the
present work, our main goal is just to analyze such a competition,
by applying TFD to non-commutative theories.

The TFD procedure is often carried out in the canonical
quantization scheme. Here we consider the thermal field
quantization by using a doubled path-integral
formalism~\cite{kha1000,seme1,seme2}, but extended to
non-commutative spaces. Although we concentrate on the $\lambda
\phi ^{4}$ theory, our procedure can be generalized to other
interacting non-commutative fields. As an application, we
calculate the two-point function at the one-loop level, with real
time, showing explicitly the effect of temperature and
non-commutativity.

The presentation is organized as follows. In Section 2 we review
some basic aspects of TFD for, in Section 3, developing  the
thermal path integral formalism for interacting fields. The
generalization to non-commutative theories is given in Section 4.
The analysis of the interplay between temperature and the
non-commutative parameter is carried out for a (1+1)-dimensional
$\lambda \phi ^{4}$  in Section 5. Our final concluding remarks
are given in Section 6.
\section{The doubling in TFD}

The duplication in TFD is defined by mapping each operator $A$
acting in the Hilbert space, $\mathcal{H}$, into another operator
$\widetilde{A}$ acting
in a Hilbert space $\widetilde{\mathcal{H}}.$ The tilde mapping $\widetilde{}%
:A\rightarrow \widetilde{A},$ called tilde conjugation rule, is
introduced from general elements of symmetry and is basically the
anti-unitary modular conjugation in $c^{\ast
}$-algebras~\cite{kha1} given by
\begin{eqnarray}
(A_{i}A_{j})\ ^{\widetilde{}} &=&\widetilde{A}_{i}\widetilde{A}_{j}, \\
(cA_{i}+A_{j})\ ^{\widetilde{}} &=&c^{\ast} \widetilde{A}_{i}+\widetilde{A}_{j}, \\
(A_{i}^{\dagger })\ ^{\widetilde{}} &=&(\widetilde{A}_{i})^{\dagger }, \\
(\widetilde{A}_{i})^{\widetilde{}} &=&A_{i}, \\
\lbrack A_{i},\widetilde{A}_{j}] &=&0.
\end{eqnarray}
Representation in the Hilbert space $\mathcal{H}_{T}=\mathcal{H\otimes }%
\widetilde{\mathcal{H}}$ is studied by considering the generator
of
symmetries as defined by%
\begin{equation}
\widehat{A}=A-\widetilde{A}.
\end{equation}%
In particular the generator of time translation is given by $\widehat{H}=H-%
\widetilde{H},$ where $H$ is the Hamiltonian of the system.

In order for a quantum field theory in the doubled Hilbert space
$\mathcal{H} _{T} $ to be constructed, we consider initially a
free boson system, and apply the tilde conjugation rules to
introduce, up to normalization factors, the following generating
functional~\cite{seme1,seme2}
\begin{eqnarray}
Z_{0}&\simeq &\int D\phi D\widetilde{\phi }e^{-iS}\nonumber \\
&=&\int D\phi D\widetilde{\phi }\exp [-i\int dx\widehat{L}],
\end{eqnarray}
where $\widehat{L}=L-\widetilde{L}$ is the doubled Lagrangian
density, given by
\begin{equation}
\widehat{L}=\frac{1}{2}\phi (\square +m)\phi -J\phi -\frac{1}{2}\widetilde{%
\phi }(\square +m)\widetilde{\phi }+\widetilde{J}\widetilde{\phi }
\end{equation}%
Such a functional can then be written as
\begin{eqnarray}
Z_{0}\simeq \exp \big\{-\frac{i}{2}\int dx dy &[J(x)(\square
+m^{2}-i\varepsilon
)^{-1}J(y)\nonumber \\
- &\widetilde{J}(x)(\square +m^{2}+i\varepsilon )^{-1}\widetilde{J}%
(y)]\big\}.
\end{eqnarray}%
From $\widehat{L},$ we write the equations of motion, such that,
the Feynman propagator for the non-tilde variables is given as
usual, $(\square +m^{2}+i\varepsilon )G_{0}(x)=-\delta (x).$ \ For
the tilde variables we have, $(-1)(\square +m^{2}-i\varepsilon
)\widetilde{G}_{0}(x)=-\delta (x),$
\ resulting that%
\begin{equation}
\widetilde{G}_{0}(x)=-G_{0}^{\ast }(x).
\end{equation}%
With these results, we find the normalized functional%
\begin{equation}
Z_{0}[\mathbf{J}^{T},\mathbf{J}]=\exp \{-\frac{i}{2}\int dxdy[\mathbf{J}%
^{T}(x)\mathbf{G}_{0}(x-y)\mathbf{J}(y)]\},
\end{equation}%
where%
\begin{equation}
\mathbf{J}(x)=\left(
\begin{array}{c}
J(x) \\
\widetilde{J}(x)%
\end{array}%
\right) ,\ \ \ \mathbf{J}^{T}(x)=(J(x),\widetilde{J}(x)),
\end{equation}%
and%
\begin{equation}
\mathbf{G}_{0}(x)=(\mathbf{G}_{0}^{ab}(x))=\left(
\begin{array}{cc}
G_{0}(x) & 0 \\
0 & -G_{0}^{\ast }(x)%
\end{array}%
\right) .
\end{equation}%
We show then that%
\begin{equation}
\mathbf{G}_{0}(x-y)=i\frac{\delta
^{2}Z[\mathbf{J}^{T},\mathbf{J}]}{\delta \mathbf{J}(y)\delta
\mathbf{J}^{T}(x)}|_{J=J^{T}=0},
\end{equation}%
where the short notation we have used is such that, for
instance,\newline $\mathbf{G}_{0}^{ab}(x-y)=i\delta
^{2}Z_{0}[\mathbf{J}^{T},\mathbf{J}]/\delta J_{a}(x)\delta
J_{b}(y)|_{J=J^{T}=0}.$ The component $\mathbf{G}_{0}^{11}(x)$
is the physical component. On the other hand, $\mathbf{G}_{0}^{22}(x)=$ $%
\widetilde{G}_{0}(x)=-G_{0}^{\ast }(x)$ is, up to the complex
conjugation and the minus sign, the physical component as well.
The role played by the doubling is that we can explore linear
mapping among the tilde and non-tilde components of the
propagator; or in an equivalent way, among the components of the
generating functional. This is the mechanism to introduce an extra
parameter in the theory, which will be used as the temperature.

\section{Bogoliubov transformation and thermal-path integral}
We proceed by introducing rotations  with the nature of a
Bogoliubov transformation in the Fourier components of the
propagator. We have
first a mapping $\mathcal{B}(\beta ):\ Z_{0}[\mathbf{J}^{T},\mathbf{J}%
]\mapsto Z_{0}[\mathbf{J}^{T},\mathbf{J;}\beta ],$ \ such that
\begin{equation}
\mathbf{G}(x-y;\beta )^{ab}=\frac{1}{(2\pi )^{4}}\int d^{4}k\,\mathbf{G}%
_{0}(k;\beta )^{ab}e^{-ik(x-y)}  \label{adre3}
\end{equation}%
where $\mathbf{G}(k;\beta )^{ab}=\mathcal{B}^{-1}(k_{0})\mathbf{G}%
_{0}(k)^{ab}\mathcal{B}(k_{0}),$ with
\begin{equation}
(\mathbf{G}_{0}(k)^{ab})=\left(
\begin{array}{cc}
\frac{1}{k^{2}-m^{2}+i\epsilon } & 0 \\
0 & \frac{-1}{k^{2}-m^{2}-i\epsilon }%
\end{array}%
\right) ,  \label{andre2}
\end{equation}
\begin{equation}
\mathcal{B}(k;\beta )=\left(
\begin{array}{cc}
\,\,\,\,\,u\,\,(k;\beta ) & -v(k;\beta ) \\
-v(k;\beta ) & \,\,u(k;\beta )%
\end{array}%
\right) ,
\end{equation}%
such that $u\,\,^{2}(k;\beta )-v\,\,^{2}(k;\beta )=1$ and
\begin{eqnarray}
u(k;\beta ) &=&\frac{1}{[1-e^{-\beta w_{k}}]^{1/2}},  \label{cos1} \\
v(k;\beta ) &=&\frac{1}{[e^{\beta w_{k}}-1]^{1/2}}.  \label{cos2}
\end{eqnarray}%
Using the definition of $\mathcal{B}(k;\beta )$, the components  $\mathbf{G}%
(k;\beta )^{ab}$ read~\cite{kha1}
\begin{eqnarray}
\mathbf{G}_{0}(k;\beta )^{11} &=&G_{0}(k)-\,n(k_{0};\beta
)\,[G_{0}(k)-G_{0}^{\ast }(k)],  \label{lei1} \\
\mathbf{G}_{0}(k;\beta )^{22} &=&-G_{0}^{\ast }(k)+n(k_{0};\beta
)\,[G_{0}(k)-G_{0}^{\ast }(k)],  \label{lei2} \\
\mathbf{G}_{0}(k;\beta )^{12} &=&\mathbf{G}_{0}(k;\beta )^{21}=-\
\,[n(k_{0};\beta )+n(k_{0})^{2}]^{1/2}\,[G_{0}(k)-G_{0}^{\ast
}(k)],  \label{lei3}
\end{eqnarray}%
where  $G_{0}(k)-G_{0}^{\ast }(k)=2\pi i\,\,\delta (k^{2}-m^{2})$
and
\begin{equation}
n(k_{0};\beta )=\frac{1}{e^{\beta w_{k}}-1}.
\end{equation}%
The generating functional is then given by
\begin{equation}
Z_{0}[\mathbf{J}^{T},\mathbf{J};\beta ]=\exp \{-\frac{i}{2}\int dxdy[\mathbf{%
J^{T}}(x)\mathbf{G}_{0}(x-y;\beta )\mathbf{J}(y)]\}.
\end{equation}
Therefore, it is simple to show that, the thermal two-point
function of the free scalar field is
\begin{equation}
\mathbf{\tau}_{0}^{ab}(x-y;\beta )=-\frac{\delta ^{2}Z[\mathbf{J}^{T},\mathbf{J}%
,\beta ]}{\delta J_{a}(x)\delta J_{b}(y)}|_{J=J^{T}=0},
\end{equation}%
reproducing in a right way the results coming from the canonical
formalism.
In order to treat interaction, we consider%
\begin{eqnarray}
\widehat{L} &=\frac{1}{2}\partial _{\mu }\phi (x)\partial ^{\mu }\phi (x)-%
\frac{m^{2}}{2}\phi ^{2}(x)+L_{int} \nonumber\\
&\quad-\frac{1}{2}\partial _{\mu }\widetilde{\phi }(x)\partial ^{\mu }\widetilde{%
\phi }(x)+\frac{m^{2}}{2}\widetilde{\phi
}^{2}(x)-\widetilde{L}_{int}.
\end{eqnarray}%
In this case, as for the zero-temperature formalism and
considering first the doubling structure, the functional
$Z[\mathbf{J}^{T},\mathbf{J}]$ fulfills the following equation
\begin{equation}
(\square +m^{2})\frac{\delta Z[\mathbf{J}^{T},\mathbf{J}]}{i\delta \mathbf{J}%
(x)}+\widehat{L}_{int}\left( \frac{1}{i}\frac{\delta }{\delta \mathbf{J}};%
\frac{1}{i}\frac{\delta }{\delta \mathbf{J}^{T}}\right) Z[\mathbf{J}^{T},%
\mathbf{J}]=\mathbf{J}(x)Z[\mathbf{J}^{T},\mathbf{J}],
\end{equation}%
with solution%
\begin{equation}
Z[\mathbf{J}^{T},\mathbf{J}]=N\exp \left[ i\int
dx\widehat{L}_{int}\left(
\frac{1}{i}\frac{\delta }{\delta \mathbf{J}};\frac{1}{i}\frac{\delta }{%
\delta \mathbf{J}^{T}}\right) \right]
Z_{0}[\mathbf{J}^{T},\mathbf{J}],
\end{equation}%
where $\widehat{L}_{int}\left( \frac{1}{i}\frac{\delta }{\delta \mathbf{J}}%
\right) =L_{int}\left( \frac{1}{i}\frac{\delta }{\delta J}\right) -%
\widetilde{L}_{int}\left( \frac{1}{i}\frac{\delta }{\delta \widetilde{J}}%
\right) .$ \ In order to introduce a  temperature-dependent
generating
functional, we map $\mathcal{B}(\beta ):Z[\mathbf{J}^{T},\mathbf{J}%
]\rightarrow Z[\mathbf{J}^{T},\mathbf{J;}\beta ],$ \ by mapping $\mathcal{B}%
(\beta ):Z_{0}[\mathbf{J}^{T},\mathbf{J}]\rightarrow Z_{0}[\mathbf{J}^{T},%
\mathbf{J;}\beta ]$, as before, resulting in
\begin{equation}
Z[\mathbf{J}^{T},\mathbf{J},\beta ]=\frac{\exp \left[ i\int dx\widehat{L}%
_{int}\left( \frac{1}{i}\frac{\delta }{\delta \mathbf{J}};\frac{1}{i}\frac{%
\delta }{\delta \mathbf{J}^{T}}\right) \right] Z_{0}[\mathbf{J}^{T},\mathbf{J%
};\beta ]}{\exp \left[ i\int dx\widehat{L}_{int}\left( \frac{1}{i}\frac{%
\delta }{\delta \mathbf{J}};\frac{1}{i}\frac{\delta }{\delta \mathbf{J}^{T}}%
\right) \right] Z_{0}[\mathbf{J}^{T},\mathbf{J};\beta ]|_{\mathbf{J=J}^{T}=0}%
}.
\end{equation}%
When  $\beta \rightarrow \infty $ \ ($T\rightarrow 0$), \ the
usual results for zero temperature are recovered.   Now we turn
our attention to the non-commutative $\lambda \phi ^{4}$ theory at
finite temperature and real time, using the generating functional
introduced above.
\section{TFD in the non-commutative plane}
We obtain a non-commutative field theory changing the usual
product between fields by the Moyal $\star $-product \cite{matt12}
in the Lagrangian density, i.e.
\begin{equation}
(\phi _{1}\star \phi _{2})(x)=e^{i\theta _{\mu \nu }\frac{\partial }{%
\partial \xi ^{\mu }}\frac{\partial }{\partial \zeta ^{\nu }}}\phi
_{1}(x+\xi )\phi _{2}(x+\zeta )\big|_{\xi =\zeta =0},
\end{equation}%
where the non-commutative parameter $\theta ^{\mu \nu }$ is skew-symmetric ($%
\theta ^{\mu \nu }=-\theta ^{\nu \mu }$). This product reproduces
a non-commutative space-time~\cite{snyder}
\begin{equation}
\lbrack x^{\mu },x^{\nu }]^{\star }=x^{\mu }\star x^{\nu }-x^{\nu
}\star x^{\mu }=2i\theta ^{\mu \nu }.
\end{equation}%
Then action of a $\phi ^{4}$-theory over this non-commutative
space-time is \cite{nikita}
\begin{equation}
S=S_{0}+S_{int}=\int d^{4}x\left[ \frac{1}{2}(\partial \phi )^{2}+\frac{1}{2}%
m^{2}\phi ^{2}+\frac{g}{4!}(\phi \star \phi \star \phi \star \phi
)(x)\right] .
\end{equation}%
With the results in the last section , we introduce the generating
functional for this non-commutative theory in the TFD formalism,
i.e.
\begin{equation}
Z[\mathbf{J}^{T},\mathbf{J},\beta ]^{\star }=\frac{\exp \left[ i\int dx%
\widehat{L}_{int}\left( \frac{1}{i}\frac{\delta }{\delta \mathbf{J}};\frac{1%
}{i}\frac{\delta }{\delta \mathbf{J}^{T}}\right) ^{\star }\right] Z_{0}[%
\mathbf{J}^{T},\mathbf{J};\beta ]}{\exp \left[ i\int dx\widehat{L}%
_{int}\left( \frac{1}{i}\frac{\delta }{\delta \mathbf{J}};\frac{1}{i}\frac{%
\delta }{\delta \mathbf{J}^{T}}\right)^{\star } \right] Z_{0}[\mathbf{J}^{T},\mathbf{J%
};\beta ]|_{\mathbf{J=J}^{T}=0}}.
\end{equation}%

Let us expand this functional, up to a normalizing factor,
\begin{eqnarray}
Z[\mathbf{J}^{T},\mathbf{J},\beta ]^{\star } &=&Z_{0}[\mathbf{J}^{T},\mathbf{%
J},\beta ]+\frac{ig^{2}}{4!}\sum_{n=1}^{2}(-1)^{n+1}a_{n}\nonumber\\
&\quad&+\frac{1}{2!}\left( \frac{ig^{2}}{4!}\right)
^{2}\sum_{n=1}^{2}\sum_{m=1}^{2}(-1)^{n+m+2}b_{nm}+\ldots ,
\label{lei9}
\end{eqnarray}%
where
\begin{eqnarray}
a_{n}\equiv \int  &&dz\left( \dfrac{\delta }{\delta
J_{n}(z)}\right)
^{4\star }Z_{0}[\mathbf{J}^{T},\mathbf{J},\beta ], \\
b_{nm}\equiv \int \int  &&dzdw\left( \dfrac{\delta }{\delta
J_{n}(z)}\right)
^{4\star }\left( \dfrac{\delta }{\delta J_{m}(w)}\right) ^{4\star }Z_{0}[%
\mathbf{J}^{T},\mathbf{J},\beta ].
\end{eqnarray}%
In order to carry out the $\star $ -product, the currents are
written in momentum space
\begin{eqnarray}
J_{a}(x) &=&\int \frac{dk}{(2\pi )^{4}}e^{(ik^{\mu }x_{\nu })}J_{a}(k), \\
\dfrac{\delta }{\delta J_{a}(x)} &=&\int \frac{dk}{2\pi
}^{4}e^{(ik^{\mu }x_{\nu })}\dfrac{\delta }{\delta J_{a}(k)}.
\end{eqnarray}%
The product in the interacting Lagrangian then gives rise to
\begin{equation}
\left( \dfrac{\delta }{\delta J_{n}(z)}\right)^{\star 4 }=\int
\prod_{i=1}^{4}\frac{dk_{i}}{(2\pi
)^{16}}e^{-\frac{i}{2}(k_{1}\theta
k_{2})}e^{-\frac{i}{2}(k_{3}\theta
k_{4})}e^{-i(\sum_{j=1}^{^{4}}k_{j}^{\mu })z_{\nu }}\dfrac{\delta
}{\delta J_{n}(k_{i})},
\end{equation}%
where $k_{i}\theta k_{j}\equiv k_{i}^{\mu }\theta _{\mu \nu
}k_{j}^{\nu }$. Therefore, we obtain
\begin{equation}
a_{n}=\int \prod_{i=1}^{4}\frac{dk_{i}}{(2\pi )^{12}}e^{-\frac{1}{2}%
\sum_{i=1}^{2}(k_{2i-1}\theta k_{2i})}\delta (\sum_{i=1}^{4}k_{i})\dfrac{%
\delta }{\delta J_{n}(k_{i})}Z_{0}[\mathbf{J}^{T},\mathbf{J},\beta
].
\end{equation}

The two-point function of this non-commutative field theory is
\begin{equation}
\mathbf{\tau}^{ab}(k_{1},k_{2};\beta )^{\star}=-\frac{\delta ^{2}Z[\mathbf{J}^{T},%
\mathbf{J},\beta ]^{\star }}{\delta J_{a}(k_{2})\delta J_{b}(k_{1})}\bigg |_{%
\mathbf{J}^{T}=\mathbf{J}=0};
\end{equation}%
such that, up to the first order in the interacting term, we have
\begin{equation}
\mathbf{\tau}_{1}^{ab}(k_{1},k_{2};\beta )^{\star}=-\frac{ig^{2}}{4!}\sum_{n=1}^{2}(-1)^{n+1}%
\frac{\delta a_{n}}{\delta J_{a}(k_{2})\delta J_{b}(k_{1})}\bigg |_{%
\mathbf{J}^{T}=\mathbf{J}=0}.
\end{equation}%
The only physical relevant part of this matrix two-point function
is the component $\mathbf{\tau}^{11}(k_{1},k_{2};\beta )^{\star}$.
However, it contains all the components of the free matrix
propagator and, therefore, the temperature effect, that is

\begin{eqnarray}
\mathbf{\tau}_{1}^{11}(k_{1},k_{2};\beta )^{\star
}=&-&\frac{g^{2}}{6}\Lambda
(k_{1},k_{2})\mathbf{G}_{0}^{11}(k_{1};\beta )\mathbf{G}_{0}^{11}(k_{2};%
\beta )\int \frac{dp}{(2\pi ^{4})}\mathbf{G}_{0}^{11}(p;\beta ) \nonumber\\
&-&\frac{g^{2}}{3}\mathbf{G}_{0}^{11}(k_{1};\beta )\mathbf{G}%
_{0}^{11}(k_{2};\beta )\int \frac{dp}{(2\pi ^{4})}\Lambda
(k_{1},p)\Lambda
(k_{2},p)\mathbf{G}_{0}^{11}(p;\beta ) \nonumber\\
&+&\frac{g^{2}}{6}\Lambda (k_{1},k_{2})\mathbf{G}_{0}^{21}(k_{1};\beta )%
\mathbf{G}_{0}^{21}(k_{2};\beta )\int \frac{dp}{(2\pi ^{4})}\mathbf{G}%
_{0}^{22}(p;\beta )\nonumber\\
&+&\frac{g^{2}}{3}\mathbf{G}_{0}^{21}(k_{1};\beta )\mathbf{G}%
_{0}^{21}(k_{2};\beta )\int \frac{dp}{(2\pi ^{4})}\Lambda
(k_{1},p)\Lambda (k_{2},p)\mathbf{G}_{0}^{22}(p;\beta ),\nonumber\\
\label{lei5}
\end{eqnarray}

where
\begin{equation}
\Lambda (k_{i},k{j})\equiv e^{-\frac{i}{2}k_{i}\theta k_{j}}
\end{equation}
and $G_{0}^{ij}$ are the components of the  TFD propagator for the
Klein-Gordon field. These results are analyzed in the following
section.

We note in the above expression for the two-point function
$\mathbf{\tau}_{1}^{11}(k_{1},k_{2};\beta
)^{\star}$ that there are two kinds of integrals (diagrams). In the first and third lines the
phase term $\Lambda (k_{i},k{j})$ is not in the integrand. This leads to planar diagrams,
which are equivalent to the usual diagrams for the commutative theory, when $\theta\to 0$. In
the second and fourth lines there are phase terms in the integrand. This leads to non-planar
diagrams. Now, non-planar diagrams are typical of non-commutative theories and leads to the
UV-IR mixing divergence. This divergence appears when one attempts to renormalize the theory in
the UV regime. This is possible only if the external momenta do not go to zero, i.e., IR
regime.

\section{$\beta,\theta$-two point function in  (1+1)-dimensions}
We find the structure of  the function
$\mathbf{\tau}_{1}^{11}(k_{1},k_{2};\beta )^{\star }$, given in
Eq.~(\ref{lei5}), by considering $\theta $ a small quantity, such
that $\Lambda (k_{i},k{j})\simeq 1-\frac{i}{2}k_{i}\theta
k_{j}.$ Since $\mathbf{G}_{0}^{11}(k_{1};\beta )$ is given by Eq.~(\ref{lei1}%
), the $T=0$ and $\theta =0$ contribution is
\begin{equation}
\mathbf{\tau}_{1}^{11}(k_{1},k_{2})=-\frac{g^{2}}{2}G_{0}(k_{1})G_{0}(k_{2})%
\int \frac{dp}{(2\pi ^{4})}G_{0}(p),
\end{equation}%
recovering the usual result. The first line in Eq.~(\ref{lei5})
provides, for instance, the
following  $\beta$ and $%
\theta $-dependent term:%
\begin{equation}
I(k_{1},k_{2};T,\theta )=\frac{g^{2}}{6}\frac{i}{2}k_{1}\theta
k_{2}n(k_{1};\beta )G_{0}(k_{1})G_{0}(k_{2})\int \frac{dp}{(2\pi ^{4})}%
G_{0}(p).  \label{lei7}
\end{equation}%
In this case, the non-commutativity, characterized by the parameter
$\theta ,$ is competing with the temperature, described by
$n(k_{1};\beta )$. Notice that for
 high enough temperature, considering $\theta $ fixed, this
terms are present in the two-point function; and so interfering in
the measurable variables, calculated from
$\mathbf{\tau}_{1}^{11}$, as the cross-section.

The interplay between temperature and the non-commutativity
parameter to the non planar part of
$\mathbf{\tau}_{1}^{11}(k_{1},k_{2})$ is better observed in the
case of 1+1 dimensions. In this limit we have
\begin{eqnarray}
\omega_{k}^{2}=k_{0}^2=\overrightarrow{k}^{2}+m^2=k_{1}^{2}+k_{2}^{2}+k_{3}^{2}+m^2\rightarrow
k_{1}^{2}+m^2,\\
k\theta p=\sum_{i,j=0}^3 k_{i}\theta^{ij}p_{j}\rightarrow
k_{0}\theta^{01}p_{1}+k_{1}\theta^{10}p_{0}=\theta(k_0p_1-k_1p_0),
\end{eqnarray}
and the non-planar part of the two-point function is
\begin{small}
\begin{eqnarray}
\mathbf{\tau}_{np}^{11}(k_{1},k_{2};\beta)^{\star
}&=&\frac{g^2}{3}\{-\mathbf{G}_{0}^{11}(k_{1};\beta )\mathbf{G}%
_{0}^{11}(k_{2};\beta )\int \frac{dp}{(2\pi ^{2})}\Lambda
(k_{1},p)\Lambda (k_{2},p)\mathbf{G}_{0}^{11}(p;\beta )\nonumber\\&&+\mathbf{G}_{0}^{21}(k_{1};\beta )\mathbf{G}%
_{0}^{21}(k_{2};\beta )\int \frac{dp}{(2\pi ^{2})}\Lambda
(k_{1},p)\Lambda (k_{2},p)\mathbf{G}_{0}^{22}(p;\beta )\}.
\end{eqnarray}
\end{small}
Now we  write $\omega_{k_1}^{2}=k^2+m^2$, $k\theta p=\theta(k_0
p_1-k_1 p_0)$ and define two dimensionless parameters $\alpha
\equiv \beta \omega_{k}$ $\gamma \equiv k \theta p $. When the
temperature is greater than the field mass ($\beta m \ll 1$) we
have $\alpha\simeq\beta k_{1}$ and
$\gamma\simeq\alpha\frac{\theta}{\beta}(p_{1}-p_{0})$.
In this limit we have
\begin{eqnarray}
\Lambda(\gamma)&=&e^{-\frac{i}{2}\gamma}\simeq 1-\frac{i}{2}\gamma,\\
n(\alpha)&=&\frac{1}{e^{\alpha}-1}\simeq \frac{1}{\alpha}.
\end{eqnarray}
The non-planar part of the two-point function is
\begin{equation}
\mathbf{\tau}_{np}^{11}(k_{1},k_{2};\beta)^{\star
}=\frac{g^2}{3}\{-A(k_1,k_2)+\frac{1}{\beta}B(k_1,k_2)+\theta
C(k_1,k_2)+\frac{\theta}{\beta}D(k_1,k_2)\}, \label{sep281}
\end{equation}
where
\begin{equation}
A(k_1,k_2)=G_0(k_1)\left[\int\frac{dp}{(2\pi)^2}G_0(p)\right]G_0(k_2),
\end{equation}
\begin{eqnarray}
B(k_1,k_2)&=&2\pi
i\bigg\{\frac{\delta(k_1^2-m^2)}{\omega_{k_1}}\left[\int\frac{dp}{(2\pi)^2}G_0(p)\right]G_0(k_2)\nonumber\\
&& +G_0(k_1)\left[\int\frac{dp}{(2\pi)^2}G_0(p)\right]\frac{\delta(k_2^2-m^2)}{\omega_{k_2}}\nonumber\\
&&
+G_0(k_1)\left[\int\frac{dp}{(2\pi)^2}\frac{\delta(p^2-m^2)}{\omega_{p}}\right]G_0(k_2)\bigg\},
\end{eqnarray}
\begin{equation}
C(k_1,k_2)=\frac{i}{2}[\omega_{k_1}+\omega_{k_2}]\bigg\{G_0(k_1)\left[\int\frac{dp}{(2\pi)^2}(p_{1}-p_{0})G_0(p)\right]G_0(k_2)\bigg\},\\
\end{equation}
\begin{eqnarray}
D(k_1,k_2)&=&\pi[\omega_{k_1}+\omega_{k_2}]\bigg\{\frac{\delta(k_1^2-m^2)}
{\omega_{k_1}}\left[\int\frac{dp}{(2\pi)^2}(p_{1}-p_{0})G_0(p)\right]G_0(k_2)\nonumber\\
&&+G_0(k_1)\left[\int\frac{dp}{(2\pi)^2}(p_{1}-p_{0})G_0(p)\right]
\frac{\delta(k_2^2-m^2)}{\omega_{k_2}}\nonumber\\
&&+G_0(k_1)\left[\int\frac{dp}{(2\pi)^2}\frac{(p_{1}-p_{0})\delta(p^2-m^2)}{\omega_{p}}\right]G_0(k_2)\bigg\}.
\end{eqnarray}
From the right-hand side of equation (\ref{sep281}), we identify the
dependence on $\theta$ and $\beta$ explicitly: in the first term,
$-A(k_1,k_2)$ is independent of $\theta$ and $\beta$ and, in
addition, this term is the value of the two-point function in the
case of $T=0$ and $\theta = 0$; the second term depends on the
temperature, only; the third on the commutative parameter; and the
fourth term depends on the interplay of commutativity and
temperature.

We note that Eq.~\ref{sep281}) shows how is the leading behavior
of these non-planar diagrams with respect to temperature and
non-commutativity. Now, these diagrams are responsible for the
existence of a singular behavior in the limit $\theta\to 0$, in
the usual case of zero temperature $\beta\to\infty$. It is clear
from this expansion that even in the case of finite temperature,
there will be no modification of the singular behavior of the
limit $\theta\to 0$.

\section{Concluding remarks}
In short, in this paper we have developed a path integral
formalism in the context of the thermofield dynamics (TFD) to
calculate the two-point function up the one-loop level of the
non-commutative $ \phi ^{4}$ theory. The main result is the nature
of contributions arising from  different terms in the propagator:
there is a term with $T=\theta=0$, where $T$ accounts for the
temperature and $\theta$ for the non-commutativity; there are
terms depending on $\theta $ only; terms depending on $T$ only;
and (mixed)
terms depending on both $T$ and $\theta ,$ as that one given in Eqs.~(\ref%
{lei7}) and (\ref{sep281}). For a fixed $\theta $ and high
temperature, the contribution of mixed terms cannot be trivially
discarded, and so it can contribute for measured quantities. A
similar analysis is still valid for high order perturbative terms
 and that can include non-equilibrium effects, since  we have
 considered   a real  (not imaginary) time approach. Beyond that,
 many important aspects   remain to be studied, as the application to  the problems of
non-commutative gauge theory and renormalization, that have been
addressed in different ways~\cite{matt12,nikita,WNC1}. For
instance, in the context of the real-time formalism, it would be
interesting to investigate the  renormalization proof for $\phi^4$
as carried out by Grosse and Wulkenhaar~\cite{GW1,GW2,GW3}.

The results derived here emerge basically from the algebraic
structure of TFD, the c*-algebra, where the tilde conjugation
rules are identified with the modular conjugation in the standard
representation. In this context, the Bogoliubov transformation,
the other basic TFD ingredient, corresponds to a linear
transformation involving the commutants of the von Neumann
algebra. It is important to emphasize that several of these
algebraic elements are also found in non-commutative theories. A
consequence is that the case $T=\theta=0$ is easily identified
since the temperature and the non-commutativity are implemented by
mappings, in a von Neumann algebra, connected to identity. This
observation points to a close connection of such formalisms that
deserves more studies.

\section*{Acknowledgements}
The Authors thanks the CNPq and CAPES (of Brazil) for financial
support. AES thanks the Department of Physics, University of
Alberta where part of this work was developed. The work of ARQ is supported in
part by CNPq under process number 307760/2009-0.

\end{document}